\documentstyle[11pt,epsfig]{article}

\newcommand{\bea}{\begin{eqnarray}}
\newcommand{\eea}{\end{eqnarray}}
\newcommand{\ba}{\begin{array}}
\newcommand{\ea}{\end{array}}

\newcommand{\vp}{{\bar p}}

\newcommand{\be}{\begin{equation}}
\newcommand{\ee}{\end{equation}}

\newcommand{\wvp}{\bar P}

\newcommand{\ms}{\mathstrut}

\begin{document}
\small
{\rm Proceedings 'V. Workshop on Nonequilibrium Physics at Short Time Scales', Rostock/Germany, April
27-30, 1998}\\[0.2cm]
\begin{center}
\large{\bf 
Creation of boson and fermion pairs in strong fields}\\[0.2cm]
\small
S. Schmidt$^1$, A.V. Prozorkevich$^2$ and S.A. Smolyansky$^2$\\[0.1cm]
\scriptsize
$^1$Fachbereich Physik, Universit\"at Rostock,
18051 Rostock, Germany\\
$^2$Physics Department, Saratov State University, Saratov, Russia

\vspace{1cm}

\normalsize
Abstract
\end{center}
We analyze a quantum kinetic equation describing  both boson and fermion pair production.
We explore the solution of the kinetic equation in its Markovian limit. The numerical study shows an enhancement (bosons) or a suppression (fermions) of the pair creation rate according to  their  quantum statistics. The modification of the time evolution of the distribution function is small but for strong fields  more pronounced than for weak fields.


\vspace{1cm}

The pre-equilibrium evolution of the quark-gluon plasma, believed to be 
created in an ultrarelativistic heavy-ion collision, is described by means of 
a transport equation that can be  used to explore dynamical properties in  
the plasma phase. 
The formation of the QGP is assumed to proceed via the creation of a strong 
chromoelectric field in the region between the two receding nuclei after the collision. The field  subsequently decays by emitting 
quark-antiquark pairs according to the nonperturbative tunneling process of 
the Schwinger mechanism. 

The process of parton pair production within the Schwinger mechanism \cite{1} has been addressed by many authors in recent years \cite{2}, with the back reaction scenario \cite{3}  also considered. While these studies have been very useful in exploring new phenomena such as plasma oscillations, the phenomenological approaches suffer from the lack of a derivation of a source term within a kinetic theory. Indeed such a consistent field theoretical treatment leads to a modified  source term providing a non-Markovian evolution of the distribution function. This result was first obtained by Rau \cite{4}
for a constant electric field. A generalized treatment allowing for a time dependent field was demonstrated in Ref. \cite{5}. A systematic numerical study of the time structure of such a source term was provided in Ref. \cite{6} for the case of the creation of boson pairs.
In Ref. \cite{5}, using a field theoretical treatment we have derived a kinetic equation characterized by the following properties: (i) it has non-Markovian character, (ii) the distribution function of the produced particles has a non-trivial momentum dependence and (iii) due to the statistical factor $[1\pm2f_\pm(\wvp,t)]$ the production is modified according to the quantum statistics of the created pairs. 
The properties of the source term itself such as the momentum dependence and the time structure have been studied in Ref. \cite{5} for a constant and a time dependent field. 
However, a numerical analysis of the evolution of the distribution function was not provided. 
In this paper we explore the solution of the kinetic equation in the Markovian limit in order to compare fermion and boson pair production. 

Using a simple  field-theoretical model  of  charged fermions
in an external, homogeneous, time-dependent field characterized by the vector potential $A_\mu=(0,0,0,A(t))$ with $A(t)=A_3(t)$ and 
the  resulting electric field 
$E(t)=E_3(t)=-{\dot A}\ms(t)=-dA(t)/dt$, we obtained \cite{5}
the kinetic equation for the distribution function $f_\pm =f_\pm(\wvp,t)$
\bea\label{10}
\frac{\partial f_\pm
}{\partial t}+eE(t)\frac{\partial f_\pm}{\partial P_\parallel(t)}=\frac{e^2
E(t)\varepsilon_\perp^2}{2\omega^2(\vp,t)}\int_{-\infty}^t dt' \frac{
E(t')}{\omega^2(\vp,t')}\bigg[1\pm2f_\pm(t')\bigg]\cos[x(t',t)],
\eea
where the upper sign (lower sign) corresponds to fermion (boson) pair creation.
We define the total energy $\omega^2(\vp,t)=\varepsilon_\perp^2+P_\parallel^2(t)$,
the transverse energy $\varepsilon_\perp^2=m^2+\vp^2_\perp$ and
$P_\parallel(t)=p_\parallel-eA(t)$.
Furthermore $x(t',t) = 2[\Theta(\vp,t)-\Theta(\vp,t')]$ denotes the difference of the dynamical phases which are defined as
\be
\label{30}
\Theta(\vp,t) = \int^t_{t_0}dt'\omega(\vp,t')\,\,.
\ee
The lower limit in Eq. (\ref{30}), $t = t_0$, describes the system in  the asymptotic field free state, $E(t_0) = 0$.  
\begin{figure}
\parbox[t]{6cm}{
\psfig{figure=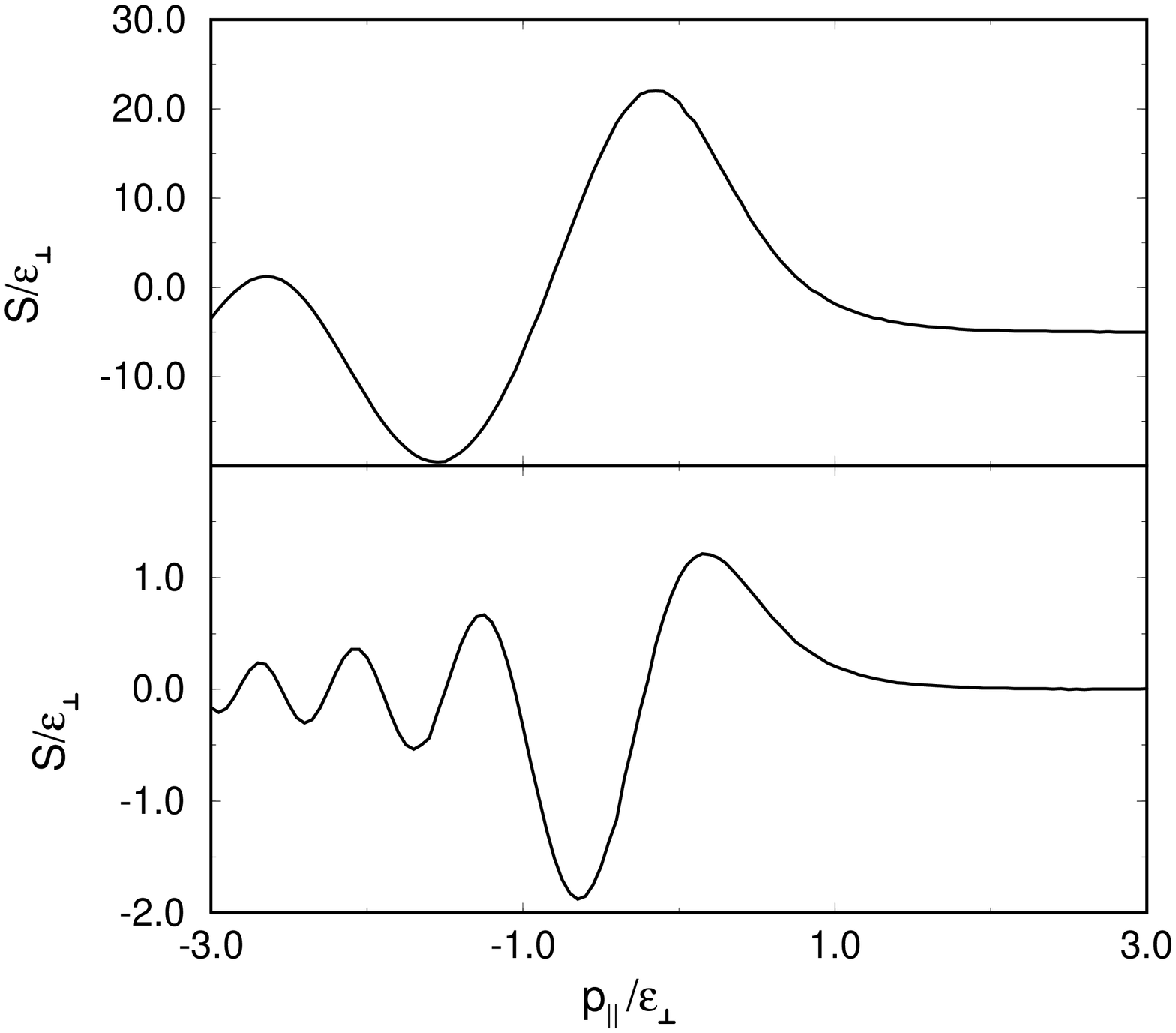,width=6cm,height=6.3cm,angle=-90}}
\hspace{1ex}
\parbox[t]{6cm}{
\psfig{figure=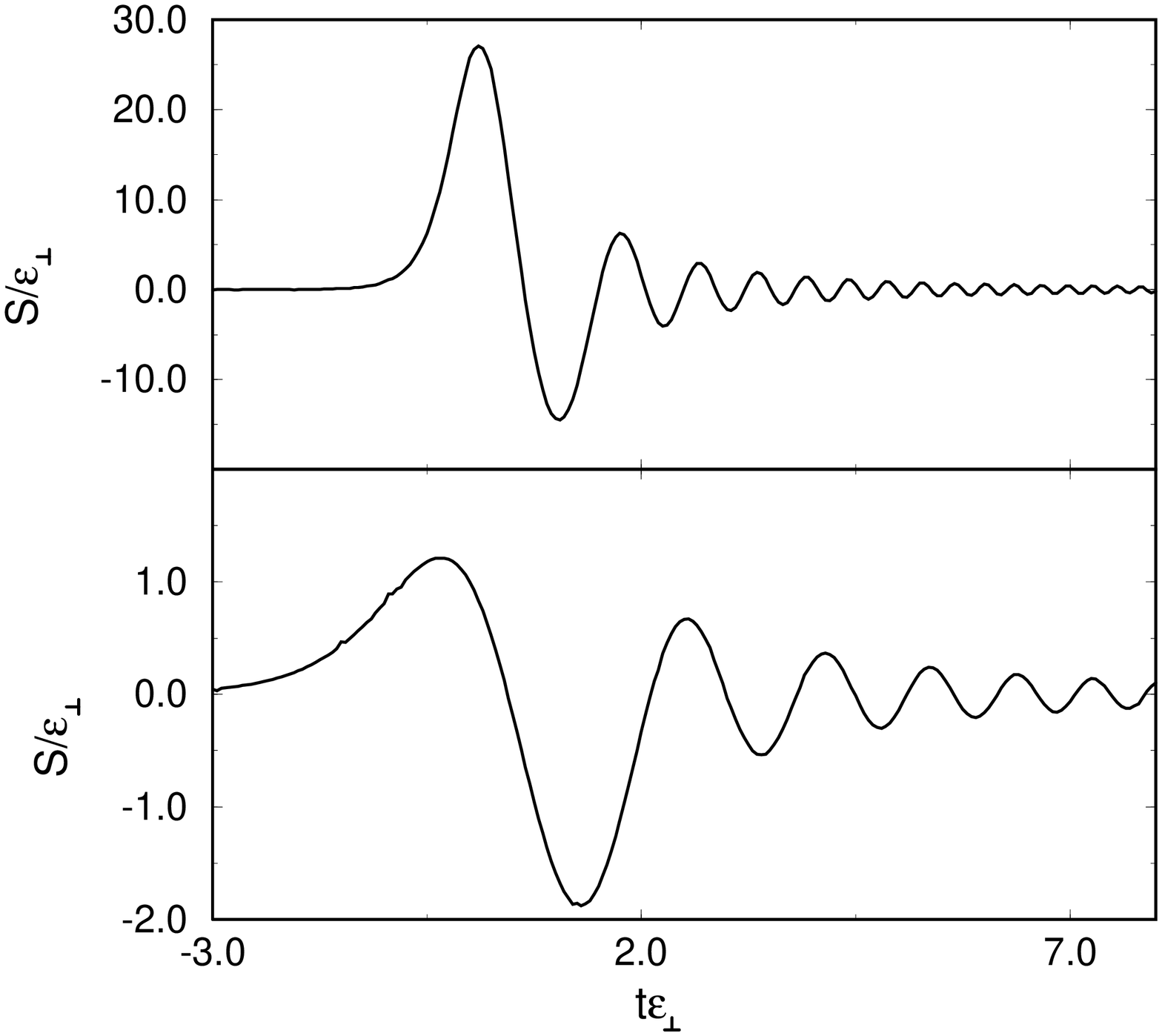,width=6cm,height=6.3cm,angle=-90}}

\vspace{0.1cm}

\caption{The dependence of the source term in Markovian limit on momentum at $t=0$ (left panels) and on time with $p_\parallel=0$ (right panels) for a strong field (upper panels), $E_0/\varepsilon_\perp^2=1.5$, and a weak field (lower panels), $E_0/\varepsilon_\perp^2=0.5$ .}
\end{figure}
The Markovian limit of Eq. (\ref{10}) is defined by the neglect of retardation effects in the source term and reads
\be\label{40}
 \frac{d\,f_\pm(\wvp ,t)}{dt}= [1\pm
2f_\pm(\wvp,t)]J(\wvp,t),   
\ee
where $J(\wvp,t)$ is the source term {\em independent} of the statistics (low density limit)
\be\label{45}
J(\wvp,t)=\frac{e^2
E(t)\varepsilon_\perp^2}{2\omega^2(\vp,t)}\int_{-\infty}^t dt' \frac{
E(t')}{\omega^2(\vp,t')}\cos[x(t',t)]\,.
\ee
Taking into account the initial condition
$\lim\limits_{t\to-\infty}f_\pm=0$, we obtain the following solution of the kinetic equation (\ref{40})
\be
\label{50}
f_\pm(\wvp ,t)=\mp\frac12 \bigg(1-\exp\bigg[\pm 2\int_{-\infty}^t dt'J(\wvp,t')\bigg]\bigg)\,.
\ee
This result is exact in the Markovian limit and holds for any time-dependent homogeneous electric field. One can show analytically that the distribution functions (\ref{50}) are positive for all times and momenta. In the lowest order of the expansion of Eq. (\ref{50}) we obtain
\be
\label{55}
f_0(\wvp ,t) = \int_{-\infty}^t dt'\frac{e^2
E(t')\varepsilon_\perp^2}{2\omega^2(\vp,t')}\int_{-\infty}^{t'} dt'' \frac{
E(t'')}{\omega^2(\vp,t'')}\cos[x(t'',t')]\,.
\ee
\begin{figure}
\parbox[t]{6cm}{
\psfig{figure=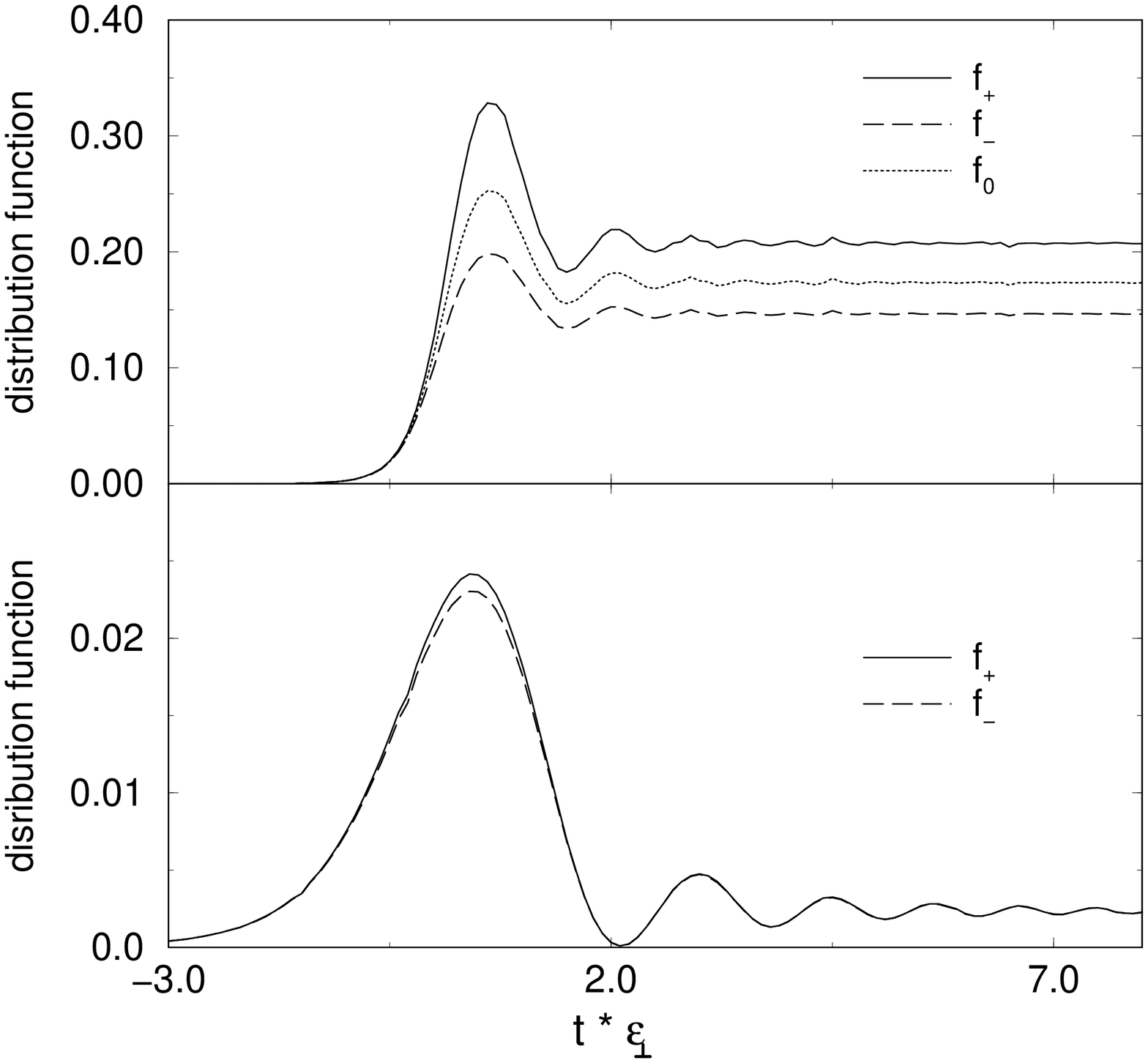,width=6cm,height=6.3cm,angle=-90}}
\hspace{3ex}
\parbox[t]{6cm}{
\psfig{figure=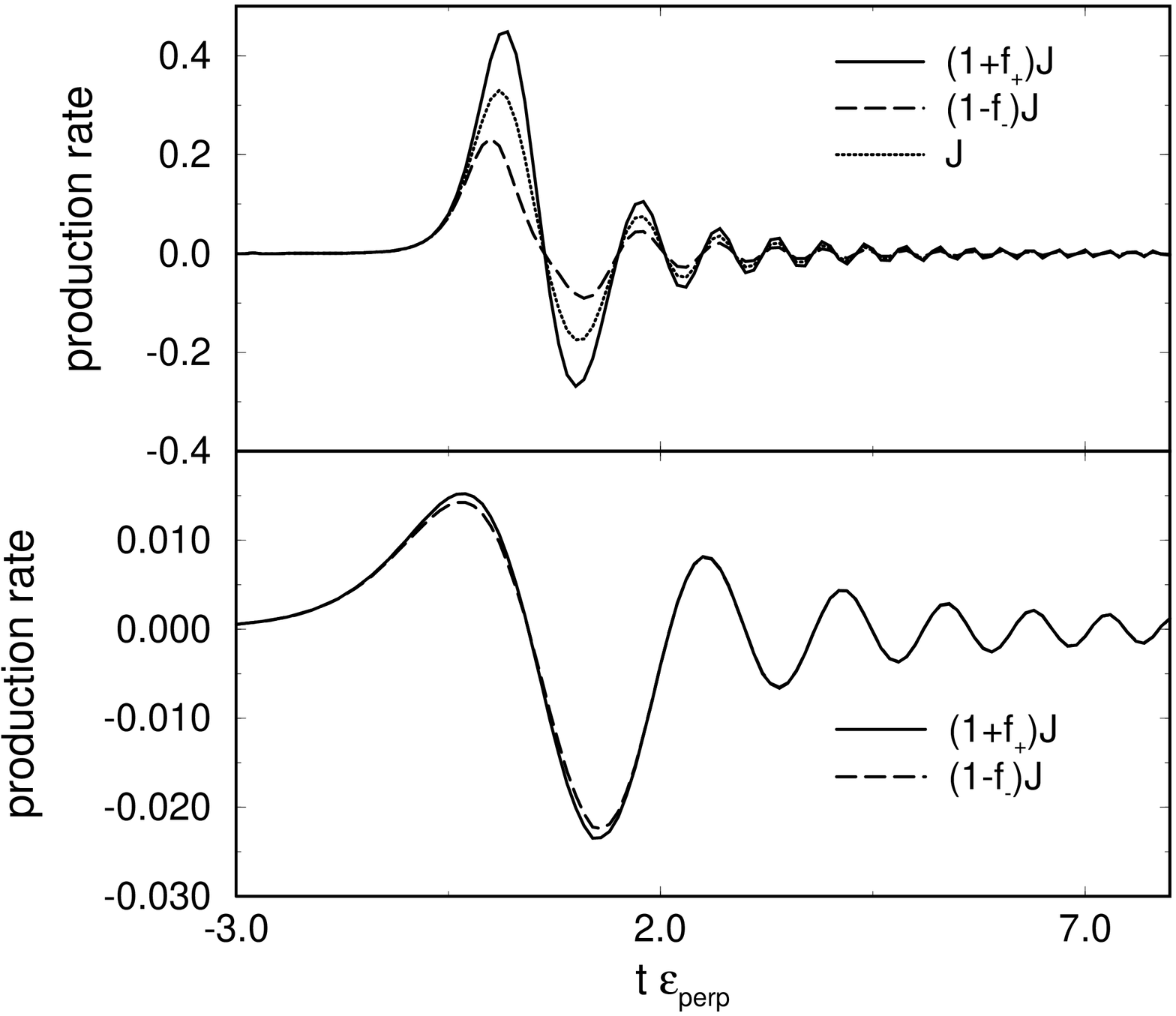,width=6cm,height=6.5cm,angle=-90}}

\vspace{0.1cm}

\caption{On the l.h.s. the time evolution of the distribution functions of bosons ($f_+$) and  fermions ($f_-$), Eq. (\ref{50}) compared with $f_0$ of Eq. (\ref{55})  at $p_\parallel=0$ for a strong field (upper panels), $E_0/\varepsilon_\perp^2=1.5$, and a weak field (lower panels), $E_0/\varepsilon_\perp^2=0.5$, is shown. On the r.h.s. the corresponding modified production rates are plotted.}
\end{figure}
For the numerical estimation we choose the simple case of  a constant electric field defined as $A(t) = t\,E_0$. 
In Fig. 1, we show the source term in Markovian limit ($J(\wvp,t)$) as function of momentum (l.h.s.) and time (r.h.s.). Resulting from the Ansatz of a constant external field, we observe oscillations which can be identified as Airy functions \cite{6}.
In the left panels of Fig. 2, the distribution functions (Eq. (\ref{50})) for fermions and bosons are plotted for a strong and a weak  electric field as function of the dimensionless time ($t\,\varepsilon_\perp$) for vanishing parallel momenta.  Compared to  $f_0$ of Eq. (\ref{55}), the distribution function for bosons ($f_+$) is enhanced while that for  fermions ($f_-$) is suppressed. For both fermions and bosons the evolution approaches a constant value at large times. For weak fields the difference between $f_+$ and $f_-$ is very small, see lower panel.

We have analytically and numerically explored the solution of a quantum kinetic equation describing particle production in the Markovian limit. We observe that the resulting production rate depends on the statistical character of the produced particles. This effect is elucidated in the right panels of Fig. 2 where the time evolution of the modified production rate is plotted. We obtain a suppression of the production rate for the creation of fermion pairs and an enhancement for boson pairs.  Although the effect is surprisingly small, it  is much more pronounced for strong fields than for weak fields.

The authors acknowledge valuable discussions with D. Blaschke, G. R\"opke, A. Schnell and V.D. Toneev. This work was supported in
part by the State Committee of Russian Federation for Higher Education under grant N 95-0-6.1-53 and by the HSP-III under the project No. 0037 6003.

\end{document}